\documentclass[11pt]{article}
\usepackage[margin=1in]{geometry}
\usepackage{amsmath,amssymb,amsthm,bm}
\usepackage{graphicx}
\usepackage{booktabs}
\usepackage[numbers]{natbib}
\usepackage[colorlinks=true,linkcolor=blue,citecolor=blue,urlcolor=blue]{hyperref}
\usepackage{microtype}
\usepackage{algorithm}
\usepackage{algpseudocode}
\usepackage{tikz}
\usetikzlibrary{positioning,arrows.meta,calc}

\newtheorem{theorem}{Theorem}
\newtheorem{assumption}{Assumption}
\newtheorem{proposition}{Proposition}
\newtheorem{corollary}{Corollary}
\theoremstyle{remark}
\newtheorem{remark}{Remark}
\newcommand{\E}{\mathbb{E}}
\newcommand{\Pn}{P_n}
\newcommand{\ind}{\mathbf{1}}

\title{Targeted Deep Survival Contrasts: Valid Inference for\\ Treatment-Specific Survival Benefit with Neural Networks}
\author{David McCoy$^{1}$ \quad Yi Li$^{1}$\\[4pt]
\small $^1$Division of Biostatistics, University of California, Berkeley}
\date{\small August 2026}

\begin{document}
\maketitle

\begin{abstract}
Neural survival models are increasingly asked to support \emph{counterfactual} claims---how much a treatment would change survival in a population---rather than only prognostic risk scores. Answering such questions from observational data requires valid inference for treatment-specific survival contrasts under confounding and covariate-dependent censoring, targets for which standard deep survival estimators are biased and provide no honest uncertainty. We propose \emph{Targeted Deep Survival Contrasts} (TDSC), which extends Targeted Deep Architectures (TDA)---targeted maximum likelihood estimation embedded in a network's weight space---to the full vector of treatment-specific survival curves $\{S_1(t), S_0(t)\}$ over a time grid, and hence to the benefit curve $S_1(t)-S_0(t)$ and the restricted mean survival time (RMST) difference. A single universal targeting path, one ridge projection of the stacked efficient influence functions onto closed-form last-layer gradients per iteration, simultaneously solves the projected estimating equations for all coordinates; a one-step residual ``top-up'' converts the plug-in into a doubly robust estimator of the unrestricted target; and a multiplier bootstrap yields simultaneous confidence bands for the benefit curve. We prove joint asymptotic linearity, band validity, and double robustness of the top-up for a cross-fitted variant requiring no Donsker conditions. Across seven Monte Carlo banks with confounded treatment, sign-varying effect heterogeneity, and dependent censoring, the TDSC plug-in attains nominal pointwise and simultaneous coverage (94.8\% and 95.0\% in the main bank) with 35\% lower MSE and narrower intervals than a per-timepoint one-step (AIPCW) built from the same nuisance fits, an advantage that persists across $n\in\{500,\dots,2000\}$, a nonlinear design, and two orders of magnitude of ridge penalty. The two estimators divide the labor cleanly: under a badly wrong outcome model the plug-in faithfully tracks its \emph{working} parameter and its intervals fail (44\% coverage), while the top-up restores nominal inference for the unrestricted causal target (94--95\%)---and in-sample diagnostics separate the two regimes replication by replication.
\end{abstract}

\section{Introduction}\label{sec:intro}

Neural networks are now standard tools for individual-level survival prediction from rich clinical inputs, including digital pathology and other high-dimensional modalities. Clinical decisions, however, hinge on \emph{contrasts}: how survival would differ under one treatment versus another. Estimating such contrasts from observational data raises two familiar obstacles---confounding and censoring that depends on treatment and covariates---and a less appreciated third: a network trained for predictive loss is a biased, inefficient estimator of any low-dimensional causal functional \citep{chernozhukov2018dml,shi2019dragonnet}, and its naive standard errors are invalid.

Semiparametric theory offers the remedy: debias a flexible initial fit by solving the efficient influence function (EIF) equation, via one-step/AIPCW corrections \citep{chernozhukov2018dml,kennedy2024review} or targeted maximum likelihood estimation (TMLE) \citep{vdl2011tl}. For survival targets these tools are mature \citep{moore2009,cai2020,rytgaard2024}, but standard implementations correct \emph{outside} the model: the one-step adds the empirical mean of the EIF to a plug-in, and classical TMLE fluctuates the output distribution after training. Both become awkward when the target is a whole \emph{vector}---two survival curves on a fine grid---where per-coordinate corrections need not respect monotonicity and can push estimates off the model space. Targeted Deep Architectures (TDA) \citep{li2025tda} takes a different route: it embeds the TMLE fluctuation \emph{inside} the network. A small targeting subset of weights (here, the final linear layers of the outcome heads) is updated along the projection of the EIF onto the per-sample gradients of the network's own training loss, so the fitted network itself becomes the debiased estimator---its outputs remain genuine survival curves, monotone by construction, and no post-hoc correction sits outside the model. TDA was demonstrated for the average treatment effect and for the marginal survival curve without treatment; the clinically decisive object---the causal contrast of treatment-specific survival curves under confounding and dependent censoring---was not addressed.

This paper closes that gap, and in doing so has to confront a question TDA's original targets did not raise: what does weight-space targeting mean when the target is $2K$ numbers rather than one? Our answer is a single \emph{universal} targeting path. Each iteration ridge-projects all $2K$ stacked EIFs onto the closed-form score gradients of the targeting layers at once, merges the per-target directions into one update, and line-searches a step; one ridge regression per iteration serves the entire curve pair. Running alongside it is a deliberately two-part inferential design. The targeted \emph{plug-in} solves the estimating equations that the network's working submodel can actually solve---the projected ones---and earns adaptive-efficiency gains from that restriction; a one-step residual \emph{top-up} then adds back whatever full-EIF bias the restriction leaves behind, recovering a regular, doubly robust estimator of the unrestricted causal target. The two estimators answer different questions, their variances are priced by different influence functions, and much of the paper is devoted to keeping that distinction honest---theoretically, and in simulations designed to make each estimator fail where it should.

\paragraph{Related work.} Neural estimators of treatment-specific survival curves exist---notably SurvITE \citep{curth2021survite}, which targets exactly our estimand with balanced discrete-time hazard networks, and counterfactual time-to-event models \citep{chapfuwa2021}---but they provide point estimates without valid confidence statements; that is the gap we fill. Conversely, survival TMLEs with simultaneous inference exist \citep{cai2020,rytgaard2024}, including universal one-dimensional least-favorable submodels that target an entire survival curve at once; but those universal paths are constructed \emph{analytically}, per estimand and likelihood. TDSC's contribution on this axis is not that a universal path exists---it is that the path is obtained \emph{generically} from the architecture's score gradients, with no estimand-specific analytic derivation, for arbitrary differentiable network backbones and stacks of pathwise-differentiable targets for which suitable influence functions are available. Causal survival forests \citep{cui2023csf} give honest inference for conditional effects at fixed horizons rather than the full population benefit curve.

\paragraph{Contributions.} We propose \emph{Targeted Deep Survival Contrasts} (TDSC), with four contributions:
\begin{enumerate}
\item the target vector $\Psi = (S_1(t_1),\dots,S_1(t_K),\allowbreak\, S_0(t_1),\dots,S_0(t_K))$ with its stacked EIFs combining treatment and censoring weights (Section~\ref{sec:setup});
\item closed-form per-sample last-layer gradients that make the universal targeting update for all $2K$ coordinates a single ridge regression per iteration (Algorithm~\ref{alg:tdsc}), with multiplier-bootstrap sup-$t$ bands for the benefit curve (Section~\ref{sec:method});
\item formal guarantees---joint asymptotic linearity, band validity, and a residual top-up that provably restores point-estimate double robustness---for a cross-fitted variant satisfying the sample-splitting premise of the underlying ADML theory, together with a design rule (and a documented failure mode) for \emph{where} the high-dimensional targeting step may be solved (Sections~\ref{sec:method}--\ref{sec:theory});
\item a simulation program---a unified main bank, nuisance-misspecification and sample-size grids, a cross-fitting comparison, and head-to-head external comparators (output-space survival TMLE, causal survival forests)---that decomposes the mechanism behind TDSC's gains and maps its operating envelope (Section~\ref{sec:sims}).
\end{enumerate}

\section{Setup, estimands, and efficient influence functions}\label{sec:setup}

\paragraph{Data.} We observe $n$ i.i.d.\ copies of $O=(X, A, \tilde T, \Delta)$: covariates $X\in\mathbb{R}^d$, binary treatment $A\in\{0,1\}$, discrete follow-up time $\tilde T\in\{1,\dots,K\}$, event indicator $\Delta$. Time is coarsened into $K$ bins; within a bin, events are recorded before censoring, and subjects surviving bin $K$ are administratively censored. Let $Y(k)=\ind\{\tilde T\ge k\}$ (at risk) and $dN(k)=\ind\{\tilde T=k,\Delta=1\}$ (event), and write $\Pn f = n^{-1}\sum_i f(O_i)$. The likelihood factorizes through the propensity $g(X)=P(A=1\mid X)$, the event hazard $h(k\mid a,x)$, and the censoring hazard $h_c(k\mid a,x)$, with conditional survival $S(t\mid a,x)=\prod_{k\le t}\{1-h(k\mid a,x)\}$ and censoring survival $S_c$ defined analogously.

\paragraph{Estimands.} Under consistency, no unmeasured confounding, coarsening-at-random censoring (censoring may depend on $A$ and $X$ but not further on the event time), and positivity for treatment and censoring, the treatment-specific curves
$S_a(t) = \E_X[S(t \mid a, X)]$, $a\in\{0,1\}$, $t\in\{1,\dots,K\}$,
are identified. Our target is the $2K$-vector $\Psi=(S_1(\cdot), S_0(\cdot))$; the \emph{benefit curve} $\beta(t)=S_1(t)-S_0(t)$ and the RMST difference $\Delta_{\mathrm{RMST}}=\sum_{t=1}^{K}\beta(t)$ (bin width one; the contrast in expected bins survived through horizon $K$) are linear in $\Psi$, so their influence functions follow by linearity. These are the quantities a treatment decision actually turns on: $\beta(t)$ is the absolute survival gain the population would realize at horizon $t$ if everyone were treated versus no one, and $\Delta_{\mathrm{RMST}}$ summarizes it as extra expected survival time. Targeting the whole vector at once, rather than one horizon at a time, is what makes simultaneous inference over the curve possible.

\paragraph{Efficient influence functions.} For each coordinate $S_a(t)$ the EIF at $P$ is the discrete-time analogue of \citet{moore2009}:
\begin{equation}\label{eq:eif}
D_{a,t}(O) = -\sum_{k=1}^{t} \frac{\ind\{A=a\}}{g_a(X)\, S_c(k-1\mid a, X)}\,\frac{S(t\mid a,X)}{S(k\mid a,X)}\big(dN(k) - Y(k)\, h(k\mid a,X)\big) + S(t\mid a,X) - S_a(t),
\end{equation}
with $g_1=g$, $g_0=1-g$. Each piece has a role worth naming. The summand is a martingale residual---the within-bin ``surprise'' of an event relative to the fitted hazard---reweighted by the inverse probability of receiving arm $a$ and remaining uncensored through bin $k-1$, so that the observed arm-$a$, still-at-risk subjects stand in for the full population; the factor $S(t\mid a,X)/S(k\mid a,X)$ propagates a hazard error at bin $k$ into its effect on survival at $t$; and the final terms center the plug-in. Stacking \eqref{eq:eif} over $a$ and $t$ gives the $n\times 2K$ influence matrix $D$ whose column means $\Pn D_{a,t}$ are the first-order bias terms of the plug-in. The targeting step of Section~\ref{sec:method} drives the \emph{projections} of these means onto the working score span to zero, and the residual top-up removes whatever the projection leaves behind.

\section{Targeted Deep Survival Contrasts}\label{sec:method}

\paragraph{Architecture and initial fit.} We use a DragonNet-style network \citep{shi2019dragonnet} adapted to discrete-time survival: a shared trunk feeding a propensity head for $g$ and two outcome heads producing hazard logits $\{h(k\mid a, x)\}_{k=1}^{K}$ for $a\in\{0,1\}$, trained on the observed-arm discrete-time negative log-likelihood (masked per-bin binary cross-entropy) plus a propensity term, with early stopping. A separate network fits the censoring hazard. Any backbone---including a frozen foundation-model encoder with a trained head---can replace the trunk. Full architecture and hyperparameters are in Appendix~\ref{app:repro}.

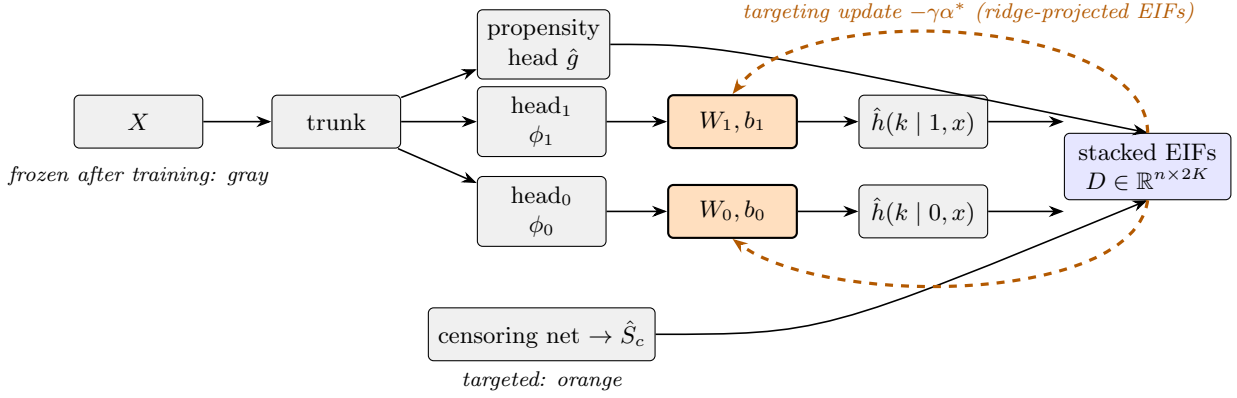
\begin{figure}[t]
\centering
\begin{tikzpicture}[
  box/.style={draw, rounded corners=2pt, minimum height=7mm, minimum width=17mm, font=\footnotesize, align=center},
  frozen/.style={box, fill=gray!12},
  targ/.style={box, fill=orange!25, thick},
  lab/.style={font=\scriptsize\itshape},
  arr/.style={-{Stealth[length=2mm]}, semithick}]
\node[frozen] (X) {$X$};
\node[frozen, right=9mm of X] (trunk) {trunk};
\node[frozen, above right=2mm and 10mm of trunk] (g) {propensity\\head $\hat g$};
\node[frozen, right=10mm of trunk] (phi1) {head$_1$\\$\phi_1$};
\node[frozen, below=2.5mm of phi1] (phi0) {head$_0$\\$\phi_0$};
\node[targ, right=8mm of phi1] (W1) {$W_1, b_1$};
\node[targ, right=8mm of phi0] (W0) {$W_0, b_0$};
\node[frozen, right=8mm of W1] (h1) {$\hat h(k\mid 1,x)$};
\node[frozen, right=8mm of W0] (h0) {$\hat h(k\mid 0,x)$};
\node[frozen, below=8mm of phi0] (cens) {censoring net $\to \hat S_c$};
\node[box, fill=blue!10, right=10mm of $(h1.east)!0.5!(h0.east)$, minimum width=22mm] (eif) {stacked EIFs\\$D \in \mathbb{R}^{n\times 2K}$};
\draw[arr] (X) -- (trunk);
\draw[arr] (trunk) -- (g);
\draw[arr] (trunk) -- (phi1);
\draw[arr] (trunk) -- (phi0);
\draw[arr] (phi1) -- (W1);
\draw[arr] (phi0) -- (W0);
\draw[arr] (W1) -- (h1);
\draw[arr] (W0) -- (h0);
\draw[arr] (h1.east) -- (eif.west|-h1.east);
\draw[arr] (h0.east) -- (eif.west|-h0.east);
\draw[arr] (g.east) .. controls +(14mm,0mm) .. (eif.north);
\draw[arr] (cens.east) .. controls +(22mm,0mm) .. (eif.south);
\draw[arr, dashed, orange!70!black, very thick]
  (eif.north) .. controls +(0,15mm) and +(10mm,13mm) .. node[above, lab, text=orange!70!black]{targeting update $-\gamma\alpha^\ast$ (ridge-projected EIFs)} (W1.north);
\draw[arr, dashed, orange!70!black, very thick]
  (eif.south) .. controls +(-2mm,-14mm) and +(12mm,-9mm) .. (W0.south);
\node[lab, below=1mm of X] {frozen after training: gray};
\node[lab, below=0mm of cens, xshift=0mm] {targeted: orange};
\end{tikzpicture}
\caption{TDSC pipeline. A DragonNet-style survival network produces per-arm hazards; the propensity head and a separate censoring network supply the weights inside the stacked EIFs. Only the final linear layers $(W_a, b_a)$ are updated, along the universal direction obtained by ridge-projecting all $2K$ EIFs onto their closed-form per-sample gradients; everything else stays frozen.}
\label{fig:pipeline}
\end{figure}

\paragraph{Targeting submodel and closed-form gradients.} Following TDA, the targeting parameters $\vartheta$ are the final linear layers of the two outcome heads, $p = 2K(F{+}1)$ parameters for head width $F$. Because those layers are linear in the head features $\phi_a(X)$ and the loss is a per-bin Bernoulli likelihood, the per-sample gradient is available in closed form:
\begin{equation}\label{eq:grad}
\frac{\partial \ell_i}{\partial W_a[k,\cdot]} = \ind\{A_i=a\}\, Y_i(k)\big(h(k\mid a, X_i) - dN_i(k)\big)\, \phi_a(X_i),
\end{equation}
and analogously for biases, giving the $n\times p$ score matrix $G$ in one vectorized pass with no per-sample automatic differentiation.

\begin{figure}[t]
\centering
\begin{tikzpicture}[scale=1.05, arr/.style={-{Stealth[length=2.2mm]}, thick}, lab/.style={font=\scriptsize}]
% score span (plane)
\draw[fill=orange!12, draw=orange!60!black]
  (-3.1,-0.9) -- (2.4,-1.5) -- (4.0,0.4) -- (-1.5,1.0) -- cycle;
\node[lab, orange!70!black] at (2.6,-1.05) {$\mathrm{span}\{\nabla_{\vartheta}\ell(O_i)\}$};
% origin
\coordinate (O) at (0,-0.25);
\fill (O) circle (1.4pt);
\node[lab, below left=0mm of O] {$\hat\vartheta$};
% EIF vector off-plane
\coordinate (D) at (1.7,1.9);
\draw[arr, blue!70!black] (O) -- (D) node[above, lab, text=blue!70!black] {$\widetilde D$ (stacked EIFs)};
% projection in-plane
\coordinate (P) at (2.1,0.12);
\draw[arr, orange!80!black, very thick] (O) -- (P) node[below right=0mm and -2mm, lab, text=orange!80!black] {$D^{\ast}_{\mathrm{proj}} = \alpha^{\ast\top}\nabla_{\vartheta}\ell$};
% residual
\draw[dashed, gray] (D) -- (P);
\node[lab, gray, rotate=-62] at (2.22,1.1) {residual $\epsilon_n$ (A\ref{a:span})};
% iterative path along plane
\draw[arr, orange!80!black, densely dotted, semithick]
  (O) .. controls (0.8,-0.15) and (1.4,0.02) .. (1.95,0.09);
\node[lab, orange!80!black] at (0.35,-1.15) {targeting steps $-\gamma\alpha^{\ast}$};
% target set
\fill[orange!80!black] (1.95,0.09) circle (1.3pt);
\node[lab] at (3.0,0.62) {$\Pn D^{w} \approx 0$};
\end{tikzpicture}
\caption{Geometry of the targeting update. The stacked nonparametric EIFs $\widetilde D$ are projected onto the span of the per-sample loss gradients of the targeting parameters; TDSC steps $\vartheta$ along the projected direction until the \emph{projected} estimating equations hold. The observable residual $\epsilon_n$ (Assumption~\ref{a:span}) measures how much of the nonparametric EIF lies outside the working score span---the gap between the two influence-function geometries and the potential efficiency gain from the restriction, not the parameter-value bias, which the oracle-bias conditions govern separately---and the full-EIF means it leaves behind are exactly what the one-step residual top-up adds back.}
\label{fig:geometry}
\end{figure}
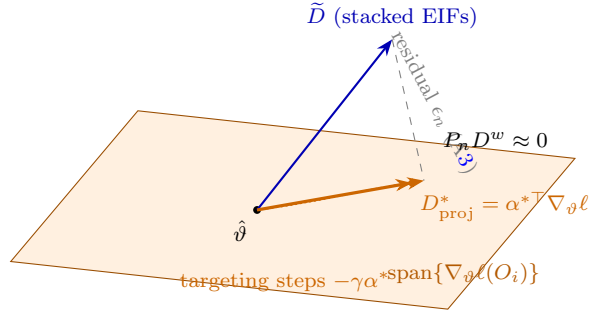

\paragraph{Universal targeting update.} Each iteration solves one ridge system for all $2K$ targets at once,
\[
\widehat\alpha_\lambda = (G^\top G + \lambda I)^{-1} G^\top D,
\]
giving a direction $\widehat\alpha_{\lambda,j}$ in weight space for each coordinate $j$ and the projected bias terms $d_j = \Pn[G\widehat\alpha_{\lambda,j}]$. The ridge is not decoration: the targeting layers have $p \approx 4{,}000$ parameters against $n=1{,}000$ observations, so the unpenalized projection is ill-posed, and $\lambda$ (scaled to the mean diagonal of $G^\top G$) is statistically active. Writing $\widehat D^{w,\lambda} = G\widehat\alpha_\lambda$ for the resulting ridge estimate of the working-submodel influence functions (Section~\ref{sec:theory} separates it from its population counterpart $D^{w}_n$), the $2K$ per-target directions are merged into one universal direction, $\alpha^* = \sum_j (d_j/\|d\|_2)\, \widehat\alpha_{\lambda,j}$ \citep[Sec.~2.3]{li2025tda}: each coordinate pulls the weights along its own direction, in proportion to how biased it currently is. The update $\vartheta \leftarrow \vartheta - \gamma\,\alpha^*$ uses a backtracking line search on $\sum_j(\Pn \widehat D^{w,\lambda}_j)^2$ ($\widehat\alpha_\lambda$ held fixed within the search); after each accepted step the hazards, influence matrix, and projection are recomputed, while $g$ and $S_c$ stay fixed.

\emph{The stopping criterion is the projected one}: iteration ends when every coordinate satisfies
\[
|\Pn \widehat D^{w,\lambda}_j| \;\le\; \widehat{\mathrm{sd}}(\widehat D^{w,\lambda}_j)\big/(\sqrt{n}\log n).
\]
These are the estimating equations of the \emph{working submodel}---the equations a restricted parameterization can actually solve. The \emph{full}-EIF means $\Pn D_j$ need not vanish inside a restricted submodel, and demanding that they do would ask the final layers for something they cannot deliver. Instead the full-EIF means are tracked, and adding them back as a one-step residual top-up, $\widehat\Psi + \Pn D$, converts the plug-in into an estimator of the unrestricted target (Proposition~\ref{prop:dr}). This division---solve what the submodel can solve, correct linearly for the rest---is the organizing idea of the method.

Two diagnostics come for free. The first is the top-up magnitude itself, $\max_j |\Pn D_j|$: how much full-EIF bias the restricted step left behind. The second is the relative projection residual $\|D - \widehat D^{w,\lambda}\|/\|D\|$, whose interpretation Section~\ref{sec:theory} makes precise: it measures how much of the nonparametric EIF lies outside the working score span---the gap between the two influence-function \emph{geometries}, hence the potential efficiency gain from the restriction---not the parameter-value bias $\Psi_n(P_0) - \Psi(P_0)$, which is governed separately by the oracle-bias conditions. Section~\ref{sec:sims} shows both diagnostics jumping sharply exactly when the working submodel is badly wrong. Finally, because the update moves weights rather than outputs, the plug-in estimator is by construction a pair of monotone survival curves.

\begin{algorithm}[t]
\caption{TDSC universal targeting (no-split form)}
\label{alg:tdsc}
\begin{algorithmic}[1]
\Require trained hazard/propensity network and censoring network; targeting parameters $\vartheta$ (final layers of the outcome heads); ridge scale $\lambda$
\State fix weights $\hat g, \hat S_c$; compute hazards $\hat h_a$, survival curves, and stacked full EIFs $D\in\mathbb{R}^{n\times2K}$ via \eqref{eq:eif}
\Loop
    \State $G \gets$ per-sample loss gradients w.r.t.\ $\vartheta$ \Comment{closed form \eqref{eq:grad}; no autograd}
    \State $\widehat\alpha_\lambda \gets (G^\top G + \lambda I)^{-1} G^\top D$; \quad $\widehat D^{w,\lambda} \gets G\widehat\alpha_\lambda$ \Comment{ridge estimate of the working EIFs}
    \State \textbf{if} $|\Pn \widehat D^{w,\lambda}_j| \le \widehat{\mathrm{sd}}(\widehat D^{w,\lambda}_j)/(\sqrt n\log n)$ for all $j$ \textbf{then break} \Comment{projected criterion}
    \State $d_j \gets \Pn \widehat D^{w,\lambda}_j$; \quad $\alpha^\ast \gets \sum_j (d_j / \lVert d\rVert_2)\, \widehat\alpha_{\lambda,j}$ \Comment{universal direction}
    \State backtracking line search: $\vartheta \gets \vartheta - \gamma\,\alpha^\ast$, accepting only if $\sum_j (\Pn \widehat D^{w,\lambda}_j)^2$ falls ($\widehat\alpha_\lambda$ fixed within the search); \textbf{break} if no $\gamma$ improves
    \State recompute $\hat h_a$, curves, and $D$ at the updated $\vartheta$
\EndLoop
\State \Return plug-ins $\widehat\Psi_j$; full-residual top-up $\widehat\Psi_j + \Pn D_j$; plug-in SEs $\widehat{\mathrm{sd}}(\widehat D^{w,\lambda}_j)/\sqrt n$, top-up SEs $\widehat{\mathrm{sd}}(D_j)/\sqrt n$; multiplier-bootstrap band; diagnostics ($\lVert D - \widehat D^{w,\lambda}\rVert/\lVert D\rVert$, $\max_j|\Pn D_j|$)
\end{algorithmic}
\end{algorithm}

\paragraph{Cross-fitted variant.} To satisfy the sample-splitting premise of the theory below without Donsker conditions on the network class, TDSC admits a $V$-fold cross-fitted form: for each fold, the nuisance networks are trained \emph{and targeted} on the complement; the held-out fold then supplies evaluation---plug-in curves, influence functions at the targeted fit, and an exact residual correction $P_{n,v}[D]$ so the EIF equations hold exactly on held-out data (by Proposition~\ref{prop:dr}, a cross-fitted one-step at the targeted nuisances). Fold contributions are averaged; the pooled per-observation influence matrix supplies variance and bands.

\paragraph{Why not target fold-wise on the held-out folds?} A tempting alternative solves the EIF equations with the full targeting submodel \emph{separately on each held-out fold}. This fold-wise high-dimensional targeting fails badly here, and the mechanism is instructive. It is \emph{not} what classical CV-TMLE \citep{zheng2011cvtmle} does: there, a single \emph{low-dimensional} fluctuation is fit by pooling the cross-validated loss over all validation folds---$O(1)$ parameters against $n$ observations---which is exactly why it is safe (the faithful CV-TMLE analogue here would pool the $2K$ one-dimensional fluctuations of Section~\ref{sec:sims}'s output-space TMLE across folds, which we have not run). TDA's submodel, by contrast, is the network's final layers ($p \approx 4{,}000$ here), so on a fold of $n/V\approx200$ observations the gradient matrix has rank at most $200$ and the fold's empirical EIF equations can be solved nearly exactly---by fitting the fold's \emph{sampling noise}, whose order $\mathrm{sd}(D)/\sqrt{n/V}$ rivals the genuine first-order bias the step is meant to remove. The resulting hazard perturbations pass through the nonlinear map $h \mapsto \prod_k(1-h(k))$, rectifying symmetric fold-level fluctuation into systematic bias, and the fold-fitted targeting layers reintroduce exactly the empirical-process term $(\Pn - P_0)(\widehat D - D_0)$ that cross-fitting is meant to eliminate---now concentrated on the smallest split. Empirically this variant gave 14-fold larger benefit-curve bias than the no-split estimator, 70.5\% pointwise and 42\% simultaneous coverage (100 replications; store preserved in the repository). The design rule: the dimension of the data-adaptive targeting step must be small relative to the split on which its equations are solved, so targeting belongs with training on the large split, and the held-out fold should receive only evaluation and the linear residual correction. Section~\ref{sec:sims} compares the no-split and cross-fitted forms empirically.

\paragraph{Inference and simultaneous bands.} Two variance prescriptions correspond to the two inferential targets of Section~\ref{sec:theory}: the \emph{top-up} estimator (unrestricted target) uses the full-EIF standard errors $\widehat{\mathrm{sd}}(D_j)/\sqrt n$, while the \emph{plug-in} (working target) uses the working-submodel standard errors $\widehat{\mathrm{sd}}(\widehat D^{w,\lambda}_j)/\sqrt n$, both at the final fit; contrasts inherit influence functions by linearity. For the benefit curve we report a sup-$t$ band over the prespecified grid: with i.i.d.\ standard normal multipliers $\xi_i$ and centered influence values $\bar D_{\beta,t} = \Pn D_{\beta,t}$, the 95\% quantile $c^*$ of $\max_t |n^{-1}\sum_i \xi_i \{D_{\beta,t}(O_i) - \bar D_{\beta,t}\}|/\widehat{\mathrm{se}}_t$ over bootstrap draws gives $\hat\beta(t) \pm c^*\widehat{\mathrm{se}}_t$ ($c^*$ floored at 1.96, a conservative choice).

\section{Theory}\label{sec:theory}

TDSC is TDA applied to a $2K$-dimensional pathwise-differentiable target, and its guarantees specialize those of \citet[Thms.~3.1--3.2]{li2025tda} and adaptive debiased machine learning (ADML) \citep{vdl2023adml}. We state the specialization for the \emph{cross-fitted} implementation (Section~\ref{sec:method}; $V$-fold, nuisances trained and targeted out-of-fold, evaluation with exact residual correction on the held-out fold), which satisfies the sample-splitting premise of those results without Donsker conditions on the network class.

Three influence-function objects must be kept distinct, because the paper's two estimators are asymptotically linear with \emph{different} ones:
\begin{itemize}
\item $D^{\mathrm{np}}_0$, the $2K$-vector of \emph{nonparametric} EIFs \eqref{eq:eif} at $P_0$: the influence function of the unrestricted target $\Psi(P_0)$, and the one the top-up estimator attains.
\item $D^{w}_n$, the \emph{working-model} EIF: the (unpenalized) $L^2(P_0)$-projection of $D^{\mathrm{np}}_0$ onto the closure of the score span of the data-adaptive working submodel $M_n\subseteq\mathcal M$ induced by partially unfreezing $\vartheta$, whose working parameter is $\Psi_n(P_0)=\Psi(\Pi_n P_0)$. Under ADML, $D^{w}_n$ converges to an \emph{oracle}-submodel EIF $D^{O}$, and the plug-in is asymptotically linear with $D^{O}$.
\item $\widehat D^{w,\lambda} = G\widehat\alpha_\lambda$, the ridge-regularized \emph{empirical} counterpart the algorithm actually computes.
\end{itemize}
Conflating the first two is what a careless reading invites: the entire variance story lives in the gap between $D^{\mathrm{np}}_0$ and $D^{O}$. For the third object, wherever $\widehat D^{w,\lambda}$ stands in for $D^{w}_n$ (the stopping rule, the plug-in standard errors) we assume $\|\widehat D^{w,\lambda} - D^{w}_n\|_{P_0} = o_p(1)$. A sufficient condition requires the ridge to vanish relative to the \emph{nonzero spectrum} of the empirical score covariance---for example, under a uniform lower bound on its nonzero eigenvalues, $\lambda_n/n \to 0$ suffices; a condition on the mean diagonal alone does not control nearly unidentified score directions. Our simulations fix the relative ridge at $0.01$ as a finite-sample stabilizer; this choice implements, but is not itself covered by, the asymptotic condition (Section~\ref{sec:sims} varies it over two orders of magnitude without material effect).

\begin{assumption}[Identification and positivity]\label{a:id}
Consistency, no unmeasured confounding, coarsening-at-random censoring; and $\delta \le g_0(X) \le 1-\delta$, $S_{c,0}(K-1\mid a, X)\ge\delta$ a.s.\ for some $\delta>0$.
\end{assumption}
\begin{assumption}[Nuisance rates]\label{a:rates}
For $f = f(k, a, x)$ write $\|f\|_{P_0}^2 = \sum_{a}\sum_{k=1}^{K} \E_{P_0} f(k,a,X)^2$. The cross-fitted weight estimators and the \emph{targeted} per-fold hazard estimators are each consistent in $\|\cdot\|_{P_0}$, respect the bounds of Assumption~\ref{a:id}, and satisfy the product rate $\|\hat h - h_0\|_{P_0}\big(\|\hat g - g_0\|_{P_0} + \|\hat S_c - S_{c,0}\|_{P_0}\big) = o_p(n^{-1/2})$ (a sufficient condition for the sharper per-$(a,t)$ sums of per-bin products appearing in the remainder).
\end{assumption}
\begin{assumption}[Gradient coverage---plug-in story only]\label{a:span}
The projection residual $\epsilon_{n,j} = D^{\mathrm{np}}_{0,j} - D^{w}_{n,j}$ stabilizes: $D^{w}_n$ converges in $L^2(P_0)$ to an oracle-model EIF $D^{O}$. \emph{This assumption is not required for Theorem~\ref{thm:main}.} It governs only the plug-in story (Proposition~\ref{prop:plugin}), and the size of $\|\epsilon_n\|$ is a dial, not a defect---but note carefully what it measures: the residual is a statement about influence-function \emph{geometry}, not parameter values. Even when the oracle submodel contains $P_0$ (so $\Psi(\Pi_0 P_0) = \Psi(P_0)$ exactly and there is no approximation bias), $D^{O}$ can remain a strict projection of $D^{\mathrm{np}}_0$ onto a smaller tangent space, with $\mathrm{Var}(D^{O}) < \mathrm{Var}(D^{\mathrm{np}}_0)$---the same parameter with a lower local efficiency bound. That is the ADML superefficiency mechanism. Approximation bias $\Psi_n(P_0) - \Psi(P_0)$ is governed separately by the oracle-bias conditions (C1)--(C2). The empirical residual $\|D - \widehat D^{w,\lambda}\|/\|D\|$ estimates the geometric gap.
\end{assumption}

\begin{theorem}[Joint asymptotic linearity and efficiency of the top-up estimator]\label{thm:main}
The full-residual top-up estimator \emph{is} the cross-fitted AIPCW one-step evaluated at the targeted nuisances (Proposition~\ref{prop:dr}); no separate ``EIF equations solved'' condition is needed. Under Assumptions~\ref{a:id} and~\ref{a:rates} alone (Assumption~\ref{a:span} not required), $\widehat\Psi^{+}$ satisfies jointly in $\mathbb{R}^{2K}$
$\widehat\Psi^{+} - \Psi(P_0) = \Pn D^{\mathrm{np}}_{0} + o_p(n^{-1/2})$, so $\sqrt n\,(\widehat\Psi^{+} - \Psi(P_0)) \rightsquigarrow N(0, \Sigma_0)$ with $\Sigma_0 = P_0 D^{\mathrm{np}}_0 (D^{\mathrm{np}}_0)^\top$: the estimator is regular and attains the nonparametric efficiency bound. Targeting enters only through Assumption~\ref{a:rates}---it must not degrade, and is designed to improve, the targeted hazards' rates---so the theorem claims no asymptotic variance advantage for weight-space targeting; any benefit is finite-sample, through better nuisances.
\end{theorem}

\begin{proposition}[Cross-fitted plug-in: working parameter and superefficiency]\label{prop:plugin}
The cross-fitted TDSC \emph{plug-in} (no top-up) is instead an ADML-type estimator of the working parameter $\Psi_n(P_0) = \Psi(\Pi_n P_0)$: under Assumption~\ref{a:span} and the conditions of \citet[Thms.~3.1--3.2]{li2025tda}---including the oracle-bias conditions (C1)--(C2) under which $\Psi_n(P_0)$ may replace $\Psi(P_0)$---it is asymptotically linear with the oracle EIF $D^{O}$, with $\mathrm{Var}(D^{O}) \preceq \Sigma_0$ in the Loewner order and strict improvement in every direction whose orthogonal residual (Assumption~\ref{a:span}) has nonzero variance (superefficiency, with the usual ADML price: nonregularity for the unrestricted target under local alternatives outside the oracle model). Its inference must therefore use the \emph{working-model} standard errors $\widehat{\mathrm{sd}}(\widehat D^{w,\lambda}_j)/\sqrt n$; pairing the plug-in with full-EIF standard errors is conservative. Both signatures---lower MSE than Theorem~\ref{thm:main}'s estimator, and over-coverage under full-EIF variance---appear in Section~\ref{sec:sims}.
\end{proposition}

\begin{corollary}[Contrasts]\label{cor:contrasts}
The benefit curve $\hat\beta(\cdot)$ and $\widehat\Delta_{\mathrm{RMST}}$ are asymptotically linear by linearity of both maps, with influence functions $D^{\mathrm{np}}_{\beta,t} = D^{\mathrm{np}}_{1,t}-D^{\mathrm{np}}_{0,t}$ and $\sum_t D^{\mathrm{np}}_{\beta,t}$ for the top-up estimator of Theorem~\ref{thm:main}, and $D^{O}_{\beta,t} = D^{O}_{1,t}-D^{O}_{0,t}$ and $\sum_t D^{O}_{\beta,t}$ for the plug-in of Proposition~\ref{prop:plugin}.
\end{corollary}

\begin{proposition}[Simultaneous band]\label{prop:band}
Under the conditions of Theorem~\ref{thm:main}, the Gaussian-multiplier estimate of the $95\%$ quantile of $\max_t |\mathbb{G}_t|/\sigma_t$, for $\mathbb{G}$ the limiting Gaussian of Corollary~\ref{cor:contrasts}, is consistent, and the sup-$t$ band covers the benefit curve over the prespecified grid with asymptotic probability at least $0.95$ (the $c^\ast\ge1.96$ floor can only enlarge it). Under Proposition~\ref{prop:plugin} plus consistent estimation of $\mathrm{Cov}(D^{O})$ by the empirical working-model IFs, the same multiplier construction applies to the plug-in.
\end{proposition}

\begin{proposition}[Top-up restores double robustness]\label{prop:dr}
Define the top-up estimator with the \emph{full} residual, $\widehat\Psi^{+} = \widehat\Psi + \Pn \widehat D$; it equals the AIPCW one-step evaluated at the targeted nuisances. Consequently, under Assumption~\ref{a:id} and convergence of the estimated nuisances to \emph{some} limits, $\widehat\Psi^{+}$ is consistent for $\Psi(P_0)$ if either the hazard limit equals $h_0$ or the weight limits equal $(g_0, S_{c,0})$ jointly---regardless of whether targeting converged.
\end{proposition}

Proof sketches are in Appendix~\ref{app:proofs}; they specialize the ADML arguments using the exact second-order remainder of the survival EIF (a telescoping identity), with cross-fitting controlling the empirical-process term.

In words: the top-up estimator is a fully classical object---efficient, regular, doubly robust for the causal target, with no assumption about how well the network's score span approximates the EIFs (Theorem~\ref{thm:main}, Proposition~\ref{prop:dr}). The plug-in is the adaptive object: it estimates the working parameter, can be strictly more efficient than the nonparametric bound allows for regular estimators, and pays for that with nonregularity when the working submodel is far from the truth (Proposition~\ref{prop:plugin}). Everything in Section~\ref{sec:sims} is an empirical test of this division of labor.

\begin{remark}[No-split implementation and scope]\label{rem:honest}
Theorem~\ref{thm:main} covers the cross-fitted top-up estimator and Proposition~\ref{prop:plugin} the cross-fitted plug-in; the no-split implementation is covered by neither as stated (it violates the sample-splitting premise of TDA's first-order expansion), and its validity at moderate $n$ is an empirical matter---Section~\ref{sec:sims} finds nominal behavior at every sample size examined ($n=500$--$2000$). Two further caveats. Double robustness of the \emph{point estimate} (Proposition~\ref{prop:dr}) does not extend to inference: consistent variance estimation still requires the relevant nuisances, and Section~\ref{sec:sims} quantifies the failure modes. Superefficiency likewise cuts both ways: narrower intervals when $M_n$ approximates $P_0$ well, and a plug-in that tracks the \emph{working} parameter---far from the truth---when it does not; the empirical price appears under outcome misspecification, where only the top-up estimators retain valid inference for $\Psi(P_0)$.
\end{remark}

\section{Simulation study}\label{sec:sims}

\paragraph{Design.} All results in this section come from one \emph{unified bank}: every estimator runs on identical datasets and, where applicable, identical nuisance fits, with the propensity always fit by a standalone network so that outcome and weight models can be blinded independently. The base (linear) data-generating process: covariates $X\sim N(0, I_{10})$; treatment $A\mid X \sim \mathrm{Bern}(\mathrm{expit}(0.9X_1 - 0.7X_2 + 0.6X_3))$, truncated to $[0.05, 0.95]$; event hazard $h(k\mid a,x) = \mathrm{expit}(-3.1 + 2k/K + 0.7x_1 + 0.5x_2 + 0.4x_3 - 0.3x_4 + a\,\tau(x))$ with $\tau(x) = -1 + 0.8x_1 + 0.6x_3$ (the treatment effect changes sign across the population); censoring hazard $\mathrm{expit}(-3.6 + 0.5x_1 + 0.4x_2 + 0.4a + 0.3x_5)$, dependent on treatment and on covariates shared with the outcome (coarsening at random holds; unadjusted estimators are biased). Two additional designs stress the network and the weights: a \emph{nonlinear} design (hazards with $\sin$, quadratic, and interaction terms; heterogeneity $\tau(x) = -1 + 0.9\tanh x_1 + 0.6 x_3 \ind\{x_2>0\}$; nonlinear censoring) and a \emph{near-positivity} design (propensity coefficients doubled, $g_0$ truncated only at $[0.02, 0.98]$). We use $K=30$ throughout; $n\in\{500, 1000, 2000\}$ for the linear design and $n=1000$ otherwise; 100--150 replications per bank (per-cell coverage MC SE $\approx$ 1.8--2.2 points; $\approx$2.8 for the 60-replication ridge banks). Ground truth is computed by Monte Carlo integration over $4\times10^5$ covariate draws (given each draw the conditional survival is evaluated exactly, so integration error is negligible at the reported precision); the true $\Delta_{\mathrm{RMST}}$ is 5.45 bins in the linear design. Tables report bias (time-averaged absolute Monte Carlo bias), the Monte Carlo SD of the estimates, the mean estimated SE (so variance-estimation behavior is diagnosable), MSE, pointwise and simultaneous coverage, and interval width.

\paragraph{Estimators.} (i) \textbf{Naive plug-in}: the initial network's curves; (ii) \textbf{unadjusted Kaplan--Meier} by arm; (iii) \textbf{IPTW$\times$IPCW KM}: weighting-only correction; (iv) \textbf{one-step (AIPCW)}: per-timepoint $\widehat\Psi_j^{\mathrm{plug}} + \Pn D_j$ at the initial fit, with its own multiplier band, plus an \emph{isotonized} variant (each arm projected onto monotone curves), which isolates how much any advantage is explained by monotonicity alone; (v) \textbf{output-space survival TMLE}, in two forms sharing the nuisances of (iv): the \emph{per-timepoint} form---each of the $2K$ coordinates gets its own one-dimensional logistic hazard fluctuation with clever covariate $H_t(k,x) = -S(t|a,x)/\{S(k|a,x)\,g_a(x)\,S_c(k{-}1|a,x)\}$, Newton-solved to convergence---and a \emph{universal} form in the spirit of \citet{cai2020,rytgaard2024}: a single one-dimensional path whose direction at each step is the $\Pn D$-weighted combination of the per-target clever covariates (computed by suffix sums), moving both arms' hazards along a shared $\epsilon$ until all $2K$ EIF equations hold jointly; (vi) \textbf{TDSC}: the plug-in with both SE types (full-EIF and working-model), the full-residual \textbf{top-up}, and the cross-fitted plug-in and top-up of Section~\ref{sec:theory}; (vii) \textbf{causal survival forests} \citep{cui2023csf}, which fit their own nuisances, evaluated as the doubly robust population benefit at horizons $\{5,10,\dots,25\}$ (these depend only on the data, which are seed-identical across banks).

\paragraph{Main results (well-specified, $n=1000$).} Table~\ref{tab:main} reports the full bank. Adjustment is essential: unadjusted KM's bias (0.123) is 27-fold the TDSC plug-in's (0.0045), and the naive plug-in carries the regularization bias debiasing exists to remove (0.025). Among the debiased estimators, the TDSC plug-in is best on every criterion: MSE 0.00102 versus 0.00158 for the one-step (a 35\% reduction on identical nuisances), pointwise coverage 94.8\% versus 93.8\%, simultaneous coverage 95.0\% versus 92.0\%, with 6\% narrower intervals (0.133 vs 0.142)---11\% narrower under the working-model SEs at 93.9\%. The MC-SD/mean-SE columns show why: the plug-in's sampling SD (0.031) sits below the full-EIF SE (0.034)---the superefficiency signature of Proposition~\ref{prop:plugin}, priced fairly by the working SE (0.032)---while the one-step's SD (0.039) slightly exceeds its SE (0.036). The full top-up gives back part of the variance advantage by construction (MSE 0.00122) in exchange for the unrestricted-target guarantee of Theorem~\ref{thm:main}.

\begin{figure}[t]
\centering
\includegraphics[width=\linewidth]{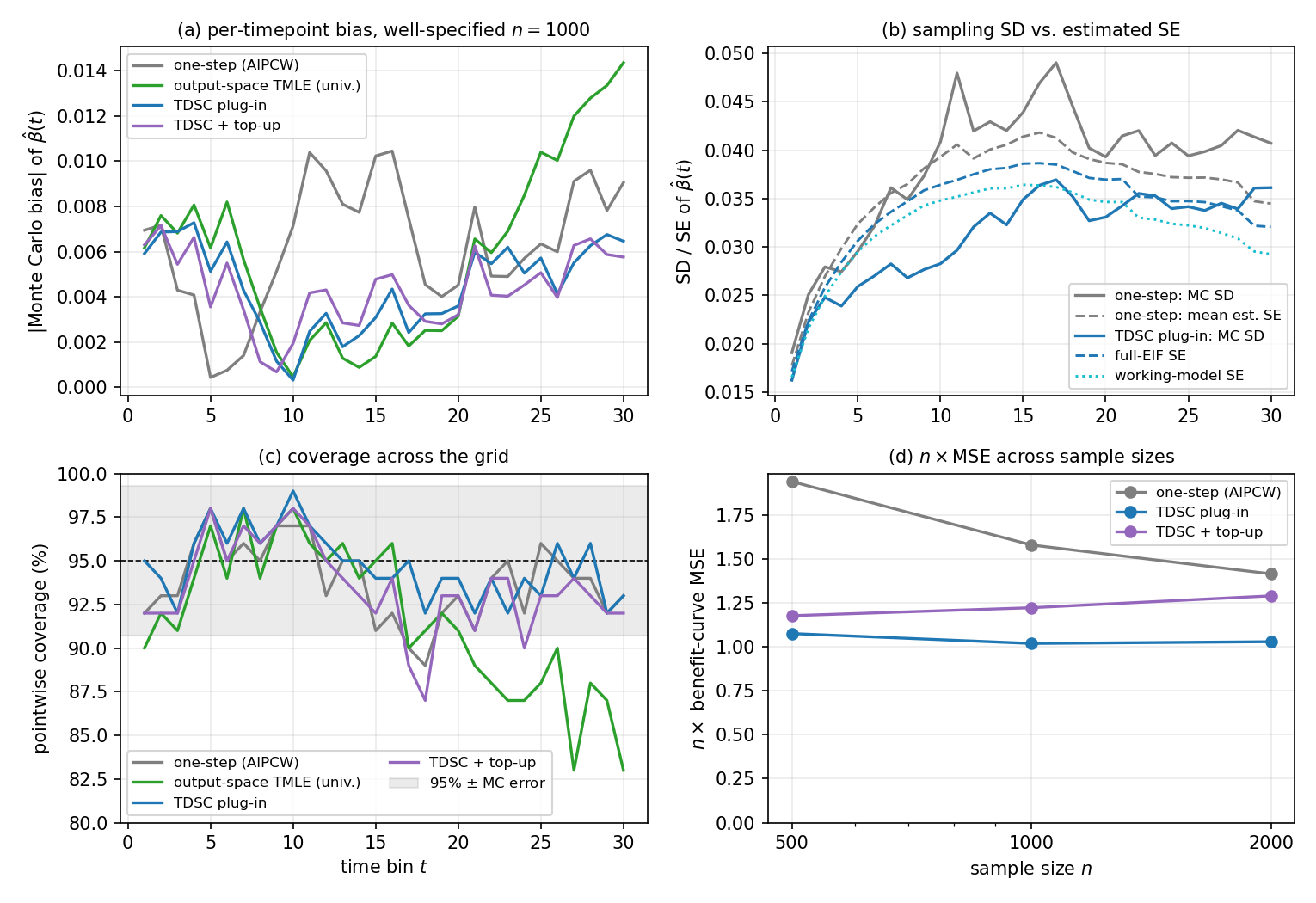}
\caption{Estimator behavior resolved across the time grid (well-specified linear design, 100 replications). (a) Per-timepoint absolute Monte Carlo bias of $\hat\beta(t)$ at $n=1000$: TDSC's advantage is roughly uniform in $t$. (b) Sampling SD versus mean estimated SE: the TDSC plug-in's SD (solid blue) runs below the one-step's (solid gray) at every horizon; the full-EIF SE (dashed) bounds it from above---the superefficiency gap of Proposition~\ref{prop:plugin}---while the working-model SE (dotted) prices it more closely. (c) Pointwise coverage across the grid with the 95\% Monte Carlo error band (shaded): TDSC stays inside the band throughout; the universal output-space TMLE degrades at late horizons. (d) $n\times$MSE across $n\in\{500,1000,2000\}$: flat near 1.0 for the TDSC plug-in, converging from above for the one-step.}
\label{fig:grid}
\end{figure}

\begin{figure}[t]
\centering
\includegraphics[width=0.8\linewidth]{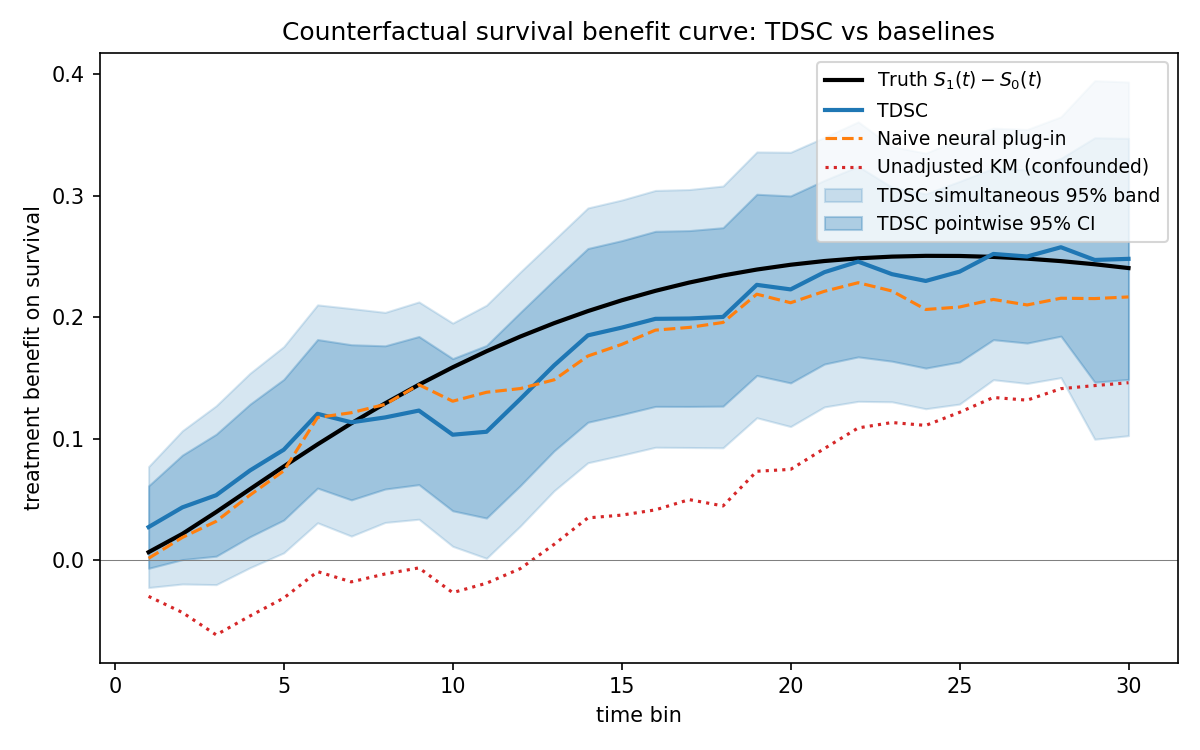}
\caption{One replication ($n=1000$, seed 7), shown for illustration. The TDSC benefit curve tracks the truth, which lies inside both the pointwise CI and the simultaneous band; the naive plug-in drifts low at later times, and unadjusted KM is severely biased (wrong-signed at early times).}
\label{fig:money}
\end{figure}

Figure~\ref{fig:grid} resolves the same comparison across the time grid, where table averages could in principle hide localized failures. They do not: TDSC's bias advantage is roughly uniform in $t$ (panel a); its sampling SD runs below the one-step's at every horizon, with the full-EIF SE bounding it from above and the working-model SE pricing it more closely, if slightly optimistically at late horizons (panel b, the superefficiency gap made visible); and its pointwise coverage stays inside the Monte Carlo error band of 95\% across the grid, while the universal output-space TMLE decays to 83--87\% at late horizons where its shared one-dimensional path is stretched thinnest (panel c). Figure~\ref{fig:money} shows one replication for illustration.

\begin{figure}[t]
\centering
\includegraphics[width=0.95\linewidth]{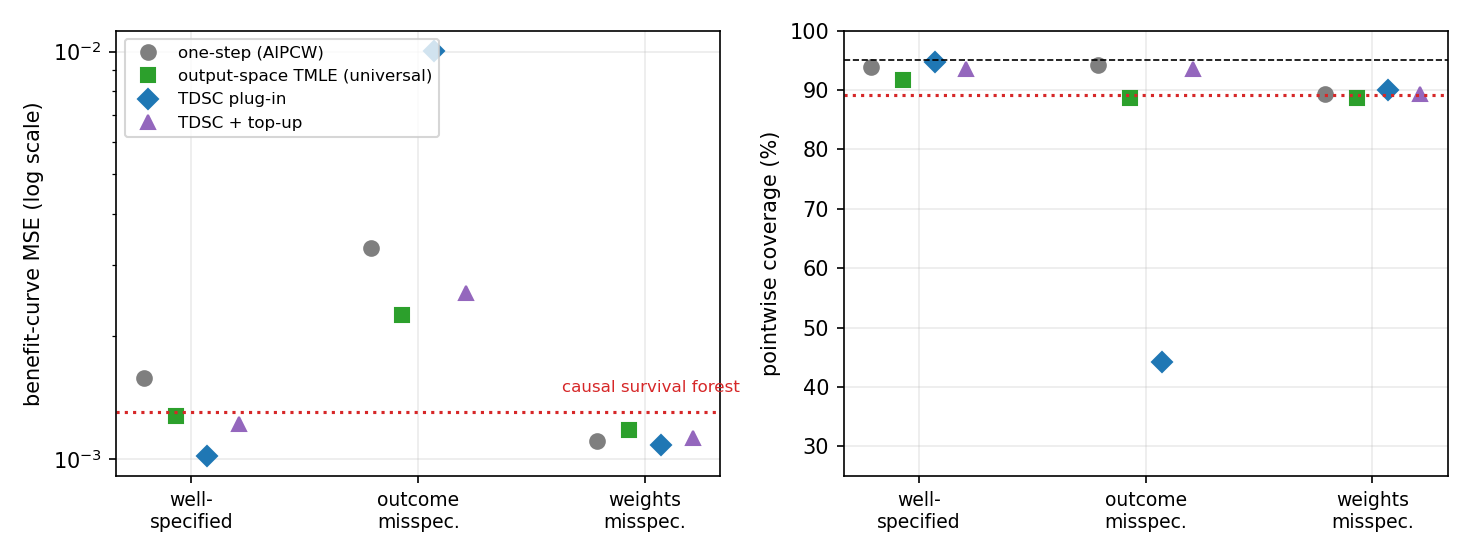}
\caption{Head-to-head comparison across nuisance scenarios (benefit-curve MSE, log scale; pointwise coverage, 95\% dashed): one-step, universal output-space TMLE, TDSC plug-in, and TDSC top-up on identical nuisances, with the causal survival forest (own nuisances, horizons 5--25) as the dotted reference. Well-specified, the estimators order one-step $\to$ output-space TMLE $\to$ TDSC; under outcome misspecification the plug-in is the most exposed---it estimates the working parameter---while the top-up and one-step remain nominal.}
\label{fig:comp}
\end{figure}

\paragraph{Mechanism decomposition.} The same-nuisance ordering in Table~\ref{tab:main} decomposes the gain along both design axes: one-step (0.00158) $\to$ output-space TMLE (0.00127 per-timepoint, 0.00128 universal) $\to$ TDSC (0.00102)---moving from additive correction to TMLE and from output space to weight space each buy roughly half the total 35\% improvement, so weight-space targeting is not redundant given TMLE-ness. The isotonized one-step is indistinguishable from the raw one-step (0.00157 vs 0.00158): monotonicity is not the mechanism. The \emph{universal} output-space path matches the per-timepoint version coordinate-for-coordinate while replacing $2K$ fluctuation sequences with one---confirming that universal output-space targeting works, as \citet{cai2020,rytgaard2024} construct analytically; TDSC's distinction is that its path falls out of the architecture's score gradients generically. Causal survival forests, fitting their own nuisances for a harder conditional-effect problem, carry visible regularization bias into this population target (bias 0.018, coverage 89.2\% at horizons 5--25); all debiased semiparametric estimators dominate here. Figure~\ref{fig:comp} summarizes.

\begin{figure}[t]
\centering
\includegraphics[width=\linewidth]{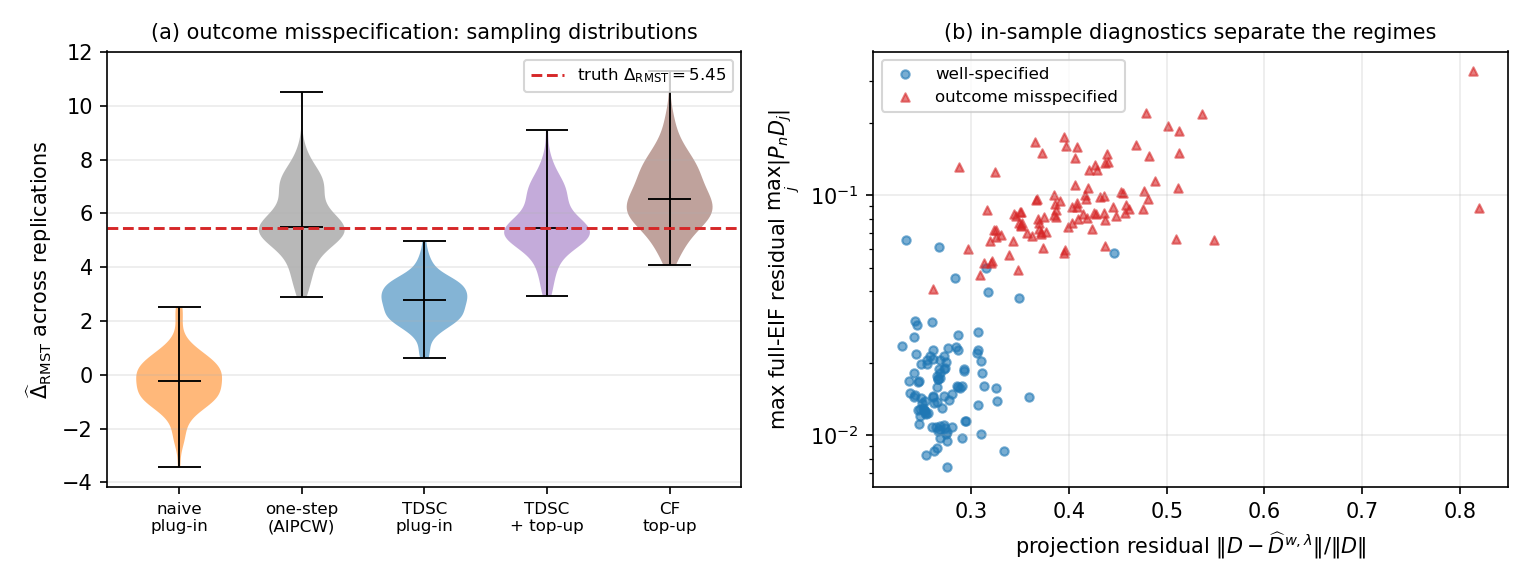}
\caption{The two-estimator story under outcome misspecification ($n=1000$, 100 replications). (a) Sampling distributions of $\widehat\Delta_{\mathrm{RMST}}$: the TDSC plug-in concentrates tightly around its \emph{working} parameter ($\approx 2.8$), far from the truth of 5.45 (dashed), while the full top-up and the cross-fitted top-up re-center on the truth; the naive plug-in is worst. (b) The two in-sample diagnostics, one point per replication: well-specified runs (circles) and outcome-misspecified runs (triangles) form nearly disjoint clusters in (projection residual, max full-EIF residual) space, so the regime is identifiable from the fitted object alone---a large top-up magnitude is the signal to report the top-up estimator.}
\label{fig:mis}
\end{figure}

\paragraph{Nuisance misspecification: the two-story framework, demonstrated.} Table~\ref{tab:mis} blinds one nuisance set at a time; Figure~\ref{fig:mis}(a) shows the resulting sampling distributions for the RMST difference. Blinding the outcome network to the main confounder and effect modifier explodes the naive plug-in (bias 0.19) and---this is the theoretically predicted part---the \emph{TDSC plug-in} (bias 0.092, coverage 44\% full-SE / 33\% working-SE): with a badly wrong working submodel, the plug-in faithfully solves the projected equations and estimates the \emph{working} parameter, which now sits far from $\Psi(P_0)$ (the oracle-bias conditions of Proposition~\ref{prop:plugin} fail). In the figure this is unmistakable: the plug-in's entire sampling distribution clusters around $\widehat\Delta_{\mathrm{RMST}} \approx 2.8$, less than a bin's width of spread but centered two and a half bins from the truth of 5.45---a precise answer to the wrong question. The \emph{full top-up} restores the unrestricted target exactly as Theorem~\ref{thm:main} says it must: bias 0.0063, coverage 93.6\%, its distribution re-centered on the truth; pairing the top-up point estimate with the one-step's initial-fit SE reaches 95.2\%. The one-step (94.2\%) and, with mild undercoverage, the output-space TMLEs ($\approx$89\%) remain stable.

Crucially, the failure is visible \emph{in-sample}, without knowing the truth. Figure~\ref{fig:mis}(b) plots the two diagnostics of Section~\ref{sec:method} for every replication: the well-specified and outcome-misspecified runs form nearly disjoint clusters, with the mean maximal full-EIF residual---exactly the top-up magnitude---jumping from 0.018 to 0.099 and the projection residual from 0.27 to 0.41. An analyst who sees a large top-up magnitude has direct evidence that the working submodel's parameter has drifted from the causal target, and should report the top-up estimator. Blinding the weight models instead leaves all EIF-corrected estimators protected through the hazard arm (biases $\approx$0.010--0.012) while weighting-only IPTW$\times$IPCW KM fails (0.076); all corrected estimators undercover mildly ($\approx$89--90\%) through the shared variance estimate---double robustness of the point estimate, not of inference.

\paragraph{Sample-size scaling.} Across $n\in\{500,1000,2000\}$ (Table~\ref{tab:nscale}; Figure~\ref{fig:grid}d) the TDSC plug-in is nominal at every size (94.7\%, 94.8\%, 95.0\% pointwise; 94.0--95.0\% simultaneous) with $n\times$MSE stable at 1.02--1.08---consistently below the one-step's 1.42--1.94, whose excess shrinks with $n$ as first-order theory predicts but is still 38\% at $n=2000$---and $\sqrt n\times$bias falling (0.18, 0.14, 0.07). The cross-fitted plug-in tracks the no-split plug-in's MSE at $n\in\{500, 1000\}$ (the sizes with cross-fit arms) but its intervals are markedly conservative (97.7--99.2\%), the predicted behavior of full- or ridge-working SEs around a superefficient point estimate whose folds are trained on $4n/5$ observations.

\paragraph{Cross-fitting.} In the well-specified bank the cross-fitted top-up---the estimator of Theorem~\ref{thm:main}---is nominal throughout (95.6\% pointwise, 97.0\% simultaneous, 96\% RMST) at a quantified price: intervals 41\% wider than the no-split plug-in (0.188 vs 0.133) and MSE 0.00243 vs 0.00122 for the no-split top-up, reflecting fold nuisances and the held-out residual's sampling noise. Under \emph{outcome misspecification}---the setting the guarantee exists for---the contrast is exactly as theory orders it: the cross-fitted \emph{plug-in} inherits the working-parameter bias (0.080, coverage 70--72\%), while the cross-fitted \emph{top-up} stays nominal (94.3\% pointwise) where the no-split plug-in fails; its point estimates are noisier than the no-split top-up's (bias 0.041 vs 0.006, width 0.234 vs 0.176), so the practical ranking under misspecification is: any full-residual top-up first, cross-fitted for the theorem-backed guarantee, no-split for finite-sample sharpness. At $n=500$ cross-fitting is disproportionately expensive (top-up MSE 0.00731, width 0.333) because each fold trains on 400 observations.

\paragraph{Robustness: nonlinear design, near-positivity, and the ridge.} On the nonlinear design the full ordering is preserved---one-step 0.00203, universal output-space TMLE 0.00158, TDSC plug-in 0.00137 at 94.8\%/94.7\% coverage---so the mechanism is not an artifact of friendly linear nuisances. Near-positivity is the stress case for everyone: with $g_0$ reaching 0.02, all estimators undercover (one-step 90.2\%, output-space TMLEs 84--86\%, TDSC plug-in 92.8\% pointwise but 87.0\% simultaneous), and the TDSC plug-in retains the best MSE (0.00143, vs 0.00373 one-step); heavy IPCW weights degrade every IF-based variance estimate, and we flag this regime as one where none of the inference here should be fully trusted without weight-stabilization. Finally, the ridge: varying the relative penalty across two orders of magnitude, $\lambda \in \{0.001, 0.003, 0.01, 0.03, 0.1\}$, moves the plug-in's MSE only within 0.00093--0.00106 and coverage within 93.9--95.3\%, while the projection residual grows from 0.18 to 0.40 exactly as shrinkage theory predicts---the ridge is statistically active in principle (Section~\ref{sec:theory}) and empirically inert over this range.

\paragraph{Convergence diagnostics, honestly.} The strict projected tolerance $\widehat{\mathrm{sd}}(\widehat D^{w,\lambda}_j)/(\sqrt n \log n)$ is formally met in only 0--13\% of replications before the 50-iteration cap binds; the projected means nonetheless end small, coverage is nominal in every well-specified configuration, and the residual top-up removes any dependence on formal convergence. The two recorded diagnostics carry real signal: under outcome misspecification the maximal full-EIF residual jumps more than five-fold ($0.018 \to 0.099$) and the projection residual by half ($0.27 \to 0.41$), giving an in-sample warning precisely where the plug-in's working target departs from the truth (Figure~\ref{fig:mis}b).

\begin{table}[t]
\centering\footnotesize
\setlength{\tabcolsep}{3.6pt}
\caption{Well-specified linear design, benefit curve ($n=1000$, $K=30$, 100 replications; coverage MC SE $\approx$ 2pp). MC SD is the Monte Carlo SD of the estimates; mean SE the average estimated standard error. RMST-difference bias/coverage in the final columns.}
\label{tab:main}
\begin{tabular}{lccccccc|cc}
\toprule
 & Bias & MC SD & mean SE & MSE & Cov. & Width & Simult. & RMST bias & cov \\
\midrule
Naive neural plug-in & 0.0252 & 0.051 & --- & 0.00352 & --- & --- & --- & $-$0.52 & --- \\
Unadjusted KM        & 0.1225 & 0.030 & --- & 0.01665 & --- & --- & --- & $-$3.68 & --- \\
IPTW$\times$IPCW KM  & 0.0132 & 0.044 & --- & 0.00218 & --- & --- & --- & $+$0.39 & --- \\
One-step (AIPCW)     & 0.0063 & 0.039 & 0.036 & 0.00158 & 93.8\% & 0.142 & 92.0\% & $-$0.14 & 94\% \\
One-step, isotonized & 0.0063 & 0.039 & --- & 0.00157 & --- & --- & --- & $-$0.14 & --- \\
Output-space TMLE (per-$t$) & 0.0042 & 0.035 & 0.034 & 0.00127 & 92.5\% & 0.134 & --- & $+$0.13 & 93\% \\
Output-space TMLE (univ.) & 0.0059 & 0.035 & 0.033 & 0.00128 & 91.7\% & 0.131 & 91.0\% & $-$0.07 & 95\% \\
TDSC plug-in (full-EIF SE) & 0.0045 & 0.031 & 0.034 & \textbf{0.00102} & \textbf{94.8\%} & 0.133 & \textbf{95.0\%} & $-$0.04 & 97\% \\
TDSC plug-in (working SE)  & 0.0045 & 0.031 & 0.032 & 0.00102 & 93.9\% & \textbf{0.126} & 94.0\% & $-$0.04 & 95\% \\
TDSC + full top-up   & 0.0044 & 0.034 & 0.034 & 0.00122 & 93.5\% & 0.133 & --- & $-$0.05 & 96\% \\
CF plug-in (working SE) & 0.0037 & 0.031 & 0.039 & 0.00098 & 97.7\% & 0.155 & --- & $-$0.03 & 97\% \\
CF top-up            & 0.0080 & 0.048 & 0.048 & 0.00243 & 95.6\% & 0.188 & 97.0\% & $-$0.22 & 96\% \\
\bottomrule
\end{tabular}
\end{table}

\begin{table}[t]
\centering\small
\caption{Nuisance-misspecification grid, benefit curve ($n=1000$, $K=30$, 100/150 replications). \emph{Outcome mis.}: outcome hazards blinded to $x_1, x_3$; \emph{weights mis.}: propensity and censoring models blinded to $x_1, x_2$. The safeguard row pairs the top-up point estimate with the one-step's initial-fit SE.}
\label{tab:mis}
\begin{tabular}{llccccc}
\toprule
Scenario & Estimator & Bias & MSE & Cov. & Width & Simult. \\
\midrule
outcome mis. & naive plug-in            & 0.1904 & 0.04028 & --- & --- & --- \\
             & one-step (AIPCW)         & 0.0107 & 0.00329 & \textbf{94.2\%} & 0.193 & 95.0\% \\
             & output-space TMLE (universal) & 0.0155 & \textbf{0.00226} & 88.7\% & 0.148 & 86.0\% \\
             & TDSC plug-in (full-EIF SE) & 0.0915 & 0.01004 & 44.2\% & 0.176 & 31.0\% \\
             & TDSC + full top-up       & \textbf{0.0063} & 0.00256 & 93.6\% & 0.176 & --- \\
             & \ \ + one-step SE (safeguard) & 0.0063 & 0.00256 & 95.2\% & 0.193 & --- \\
             & CF plug-in (working SE)  & 0.0802 & 0.00774 & 70.0\% & 0.217 & --- \\
             & CF top-up                & 0.0410 & 0.00587 & 94.3\% & 0.234 & 98.0\% \\
\midrule
weights mis. & IPTW$\times$IPCW KM      & 0.0760 & 0.00720 & --- & --- & --- \\
             & one-step (AIPCW)         & 0.0114 & 0.00111 & 89.4\% & 0.111 & 84.7\% \\
             & output-space TMLE (universal) & 0.0116 & 0.00118 & 88.7\% & 0.110 & 84.0\% \\
             & TDSC plug-in (full-EIF SE) & \textbf{0.0096} & \textbf{0.00108} & \textbf{90.1\%} & 0.110 & 87.3\% \\
             & TDSC + full top-up       & 0.0103 & 0.00113 & 89.3\% & 0.110 & --- \\
\bottomrule
\end{tabular}
\end{table}

\begin{table}[t]
\centering\small
\caption{Sample-size scaling and robustness designs, benefit curve ($K=30$). Top: linear design across $n$. Bottom: nonlinear and near-positivity designs at $n=1000$ (MSE with pointwise coverage in parentheses; TDSC simultaneous coverage in the final column), and the ridge-sensitivity range for the TDSC plug-in ($n=1000$, 60 replications per $\lambda$).}
\label{tab:nscale}
\begin{tabular}{rlcccc}
\toprule
$n$ & Estimator & $n\times$MSE & Cov. & Width & Simult. \\
\midrule
 500 & one-step      & 1.94 & 94.0\% & 0.218 & 91.0\% \\
     & TDSC plug-in  & \textbf{1.08} & \textbf{94.7\%} & 0.191 & \textbf{94.0\%} \\
     & TDSC + top-up & 1.18 & 94.0\% & 0.191 & --- \\
1000 & one-step      & 1.58 & 93.8\% & 0.142 & 92.0\% \\
     & TDSC plug-in  & \textbf{1.02} & \textbf{94.8\%} & 0.133 & \textbf{95.0\%} \\
     & TDSC + top-up & 1.22 & 93.5\% & 0.133 & --- \\
2000 & one-step      & 1.42 & 92.7\% & 0.093 & 90.0\% \\
     & TDSC plug-in  & \textbf{1.03} & \textbf{95.0\%} & 0.091 & \textbf{94.0\%} \\
     & TDSC + top-up & 1.29 & 92.6\% & 0.091 & --- \\
\bottomrule
\end{tabular}

\medskip

\begin{tabular}{lcccc}
\toprule
Design ($n{=}1000$) & one-step & univ.\ osTMLE & TDSC plug-in & TDSC simult. \\
\midrule
Nonlinear       & 0.00203 (93.2\%) & 0.00158 (92.3\%) & \textbf{0.00137} (\textbf{94.8\%}) & 94.7\% \\
Near-positivity & 0.00373 (90.2\%) & 0.00245 (84.1\%) & \textbf{0.00143} (92.8\%) & 87.0\% \\
\midrule
\multicolumn{5}{l}{\emph{Ridge sensitivity}, TDSC plug-in, $\lambda\in\{0.001,\dots,0.1\}$: MSE 0.00093--0.00106;}\\
\multicolumn{5}{l}{\quad coverage 93.9--95.3\%; projection residual 0.18$\to$0.40}\\
\bottomrule
\end{tabular}
\end{table}

\section{Discussion}\label{sec:disc}

TDSC turns a two-headed neural survival network into a valid estimator of population treatment benefit at the cost of a few ridge regressions after training, and the simulation program gives its two estimators an honest division of labor. The \emph{plug-in} is the adaptive-efficiency instrument: nominal coverage with the best MSE at every sample size examined, on nonlinear nuisances, and across two orders of magnitude of ridge penalty---but it is inference for the \emph{working} parameter, and under a badly wrong outcome model it tracks that parameter faithfully right past the truth. The \emph{full top-up} is the insurance instrument: a provably regular, doubly robust estimator of the unrestricted target that stayed nominal under outcome misspecification---the setting that breaks the plug-in---and matched the other corrected estimators' mild undercoverage under weight misspecification (a shared variance-estimation limit, not a point-estimate failure). When the theorem matters, cross-fit it; when finite-sample sharpness matters, the no-split form was nominal at every configuration we examined.

The choice between the two need not be blind. The in-sample diagnostics moved sharply in exactly the settings where the plug-in failed---top-up magnitude five-fold, projection residual by half, with nearly disjoint per-replication clusters (Figure~\ref{fig:mis}b)---so a practical protocol suggests itself: fit both, and let the top-up magnitude arbitrate. When it is small, the plug-in's working parameter is close to the causal target and its sharper intervals are trustworthy; when it is large, report the top-up.

Limitations and next steps. Near-positivity remains the shared hard regime---every IF-based variance we examined degraded there---and weight stabilization for the targeting step is open. We have not yet mapped sensitivity to the targeting-layer choice, weight-truncation levels, or the grid size $K$. And two extensions are natural: \emph{calibration-for-benefit}---targeting arm-specific curves within strata of the network's own predicted benefit, the estimand used to validate individualized-benefit models against randomized trials---and application on frozen pretrained encoders (e.g.\ pathology foundation models), which the last-layer construction supports directly.

\clearpage
\appendix
\section{Reproducibility}\label{app:repro}
All code, raw Monte Carlo stores, and the exact scripts behind every table are at \url{https://github.com/blind-contours/tdsc}. Networks: trunk $d\to64\to64$ (ELU); propensity head $64\to64\to1$; per-arm outcome heads $64\to64\to K$; censoring network $(d{+}1)\to64\to64\to(K{-}1)$. Training: full-batch Adam, lr $10^{-3}$, max 500 epochs, early stopping (patience 25) on a 20\% validation split, propensity loss weight 1.0. Targeting: ridge scale $\lambda = 0.01\times\mathrm{mean\,diag}(G^\top G)$ (a finite-sample stabilizer; the theory requires only that the relative ridge vanish, Section~\ref{sec:theory}); $\gamma_0=1$ with up to 12 halvings; max 50 iterations; stopping tolerance $\widehat{\mathrm{sd}}(\widehat D^{w,\lambda}_j)/(\sqrt n \log n)$ on the projected means. Truncation: $\hat g\in[0.05,0.95]$, $\hat S_c \ge 0.05$, conditional survival floored at $10^{-3}$ inside \eqref{eq:eif}. Bands: 1000 Gaussian-multiplier draws. Cross-fitting: $V=5$, fold networks trained and targeted on complements with the same hyperparameters. Output-space TMLE, per-timepoint form: per-coordinate logistic fluctuations solved by Newton steps ($\epsilon$ clipped to $[-5,5]$), recomputing the clever covariate each round, up to 25 rounds per coordinate; universal form: shared one-dimensional path along the $\Pn D$-weighted suffix-sum clever covariate, $\epsilon$ steps clipped to $[-0.5, 0.5]$, up to 200 steps (70--105 observed), same $\mathrm{sd}/(\sqrt n\log n)$ tolerance. Causal survival forests: \texttt{grf} 2.6.1, \texttt{causal\_survival\_forest} with \texttt{target = "survival.probability"}, 1{,}000 trees, default tuning, one forest per horizon $\{5,10,15,20,25\}$, population effect and SE from \texttt{average\_treatment\_effect}. Every bank uses a standalone propensity network and paired data seeds; replications per bank: 100 (well $n{=}1000$, outcome-misspecified, near-positivity, $n{=}500$, $n{=}2000$), 150 (weights-misspecified, nonlinear), 60 per ridge value. Truth: Monte Carlo integration over $4\times10^5$ covariate draws.

\section{Proof sketches}\label{app:proofs}

\paragraph{Theorem~\ref{thm:main}.} All influence functions in this proof are the nonparametric EIFs \eqref{eq:eif}; superscripts are omitted. Per coordinate, since $\widehat\Psi^{+}_j = \Psi_j(\hat P) + \Pn\widehat D_j$ and the EIF has the form $\widehat D_j = \widehat D_j^{\mathrm{mart}} + \hat S_j - \Psi_j(\hat P)$, the standard one-step decomposition holds exactly:
\[
\widehat\Psi^{+}_j - \Psi_j(P_0) \;=\; \Pn D_{0,j} \;+\; (\Pn - P_0)\big(\widehat D_j - D_{0,j}\big) \;+\; R_{2,j},
\qquad R_{2,j} = P_0 \widehat D_j + \Psi_j(\hat P) - \Psi_j(P_0).
\]
For the remainder, the survival EIF is exact to second order: writing $S_{h}(k)=\prod_{j\le k}(1-h(j))$ for the conditional survival under a hazard $h$, the telescoping identity
$S_{\hat h}(t) - S_{h_0}(t) = \sum_{k\le t} S_{h_0}(k-1)\,\big(h_0(k)-\hat h(k)\big)\,S_{\hat h}(t)/S_{\hat h}(k)$
gives $R_{2,j}$ as a sum over $k\le t$ of integrals of $\{\hat h(k) - h_0(k)\}$ against $\{\hat g\hat S_c(k-1) - g_0 S_{c,0}(k-1)\}/\{\hat g \hat S_c(k-1)\}$-type differences; under the positivity bounds of Assumption~\ref{a:id}, Cauchy--Schwarz gives $|R_{2,j}| \lesssim \|\hat h - h_0\|_{P_0}(\|\hat g - g_0\|_{P_0} + \|\hat S_c - S_{c,0}\|_{P_0}) = o_p(n^{-1/2})$ by Assumption~\ref{a:rates}, applied to the targeted per-fold hazards (this display also proves Proposition~\ref{prop:dr}: $R_2$ vanishes if either factor's limit is correct). For the empirical-process term, cross-fitting makes $\widehat D_j$ independent of the fold on which $(\Pn-P_0)$ is evaluated, so conditional on the training folds, $(\Pn - P_0)(\widehat D_j - D_{0,j}) = o_p(n^{-1/2})$ whenever $\|\widehat D_j - D_{0,j}\|_{P_0} = o_p(1)$, which follows from the individual nuisance consistency in Assumption~\ref{a:rates} (targeting, which moves $\vartheta$ along the score span, is designed to improve rather than degrade those rates). Summing folds yields the coordinate-wise expansion; joint convergence follows from the multivariate CLT applied to the stacked $D_0$. Proposition~\ref{prop:plugin} is \citet[Thms.~3.1--3.2]{li2025tda} applied verbatim to the cross-fitted plug-in, with superefficiency from their Theorem~3.2 under (C1)--(C2).

\paragraph{Corollary~\ref{cor:contrasts} and Proposition~\ref{prop:band}.} Both maps are bounded linear in $\Psi$, so asymptotic linearity transfers with the transformed influence functions. Given joint normality with consistently estimated $\Sigma_0$ (IF second moments), the conditional multiplier process $n^{-1/2}\sum_i \xi_i \widehat D_{\beta,t}(O_i)$ converges weakly to the same Gaussian limit (conditional multiplier CLT), so its sup-quantile is consistent for the sup-quantile of the limit; the $c^*\ge1.96$ floor only enlarges the band.

\paragraph{Proposition~\ref{prop:dr}.} By definition $\widehat\Psi^{+}_j = \Pn[\hat S(t\mid a, X)] + \Pn \widehat D_j$, which is precisely the AIPCW one-step at nuisances $(\hat h^{\mathrm{targ}}, \hat g, \hat S_c)$. Its population bias is the $R_2$ term above evaluated at those limits, which vanishes under either-arm consistency. \hfill$\square$

\bibliographystyle{plainnat}

\begin{thebibliography}{12}

\bibitem[Cai and van~der Laan(2020)]{cai2020}
W.~Cai and M.~J. van~der Laan.
\newblock One-step targeted maximum likelihood estimation for time-to-event outcomes.
\newblock \emph{Biometrics}, 76(3):722--733, 2020.
\newblock \url{https://doi.org/10.1111/biom.13172}.

\bibitem[Chapfuwa et~al.(2021)]{chapfuwa2021}
P.~Chapfuwa, S.~Assaad, S.~Zeng, M.~J. Pencina, L.~Carin, and R.~Henao.
\newblock Enabling counterfactual survival analysis with balanced representations.
\newblock In \emph{Proceedings of the Conference on Health, Inference, and Learning (CHIL)}, 2021.
\newblock \url{https://doi.org/10.1145/3450439.3451875}.

\bibitem[Chernozhukov et~al.(2018)]{chernozhukov2018dml}
V.~Chernozhukov, D.~Chetverikov, M.~Demirer, E.~Duflo, C.~Hansen, W.~Newey, and J.~Robins.
\newblock Double/debiased machine learning for treatment and structural parameters.
\newblock \emph{The Econometrics Journal}, 21(1):C1--C68, 2018.
\newblock \url{https://doi.org/10.1111/ectj.12097}.

\bibitem[Cui et~al.(2023)]{cui2023csf}
Y.~Cui, M.~R. Kosorok, E.~Sverdrup, S.~Wager, and R.~Zhu.
\newblock Estimating heterogeneous treatment effects with right-censored data via causal survival forests.
\newblock \emph{Journal of the Royal Statistical Society: Series B}, 85(2):179--211, 2023.
\newblock \url{https://doi.org/10.1093/jrsssb/qkac001}.

\bibitem[Curth et~al.(2021)]{curth2021survite}
A.~Curth, C.~Lee, and M.~van~der Schaar.
\newblock SurvITE: Learning heterogeneous treatment effects from time-to-event data.
\newblock \emph{Advances in Neural Information Processing Systems}, 34, 2021.
\newblock \url{https://proceedings.neurips.cc/paper/2021/hash/e0eacd983971634327ae1819ea8b6214-Abstract.html}.

\bibitem[Kennedy(2024)]{kennedy2024review}
E.~H. Kennedy.
\newblock Semiparametric doubly robust targeted double machine learning: a review.
\newblock In \emph{Handbook of Statistical Methods for Precision Medicine}. Chapman \& Hall/CRC, 2024.
\newblock \url{https://arxiv.org/abs/2203.06469}.

\bibitem[Li et~al.(2025)]{li2025tda}
Y.~Li, D.~McCoy, N.~Gunter, K.~Lee, A.~Schuler, and M.~J. van~der Laan.
\newblock Targeted deep architectures: A TMLE-based framework for robust causal inference in neural networks.
\newblock \emph{arXiv preprint arXiv:2507.12435}, 2025.
\newblock \url{https://arxiv.org/abs/2507.12435}.

\bibitem[Moore and van~der Laan(2009)]{moore2009}
K.~L. Moore and M.~J. van~der Laan.
\newblock Increasing power in randomized trials with right censored outcomes through covariate adjustment.
\newblock \emph{Journal of Biopharmaceutical Statistics}, 19(6):1099--1131, 2009.
\newblock \url{https://doi.org/10.1080/10543400903243017}.

\bibitem[Rytgaard and van~der Laan(2024)]{rytgaard2024}
H.~C.~W. Rytgaard and M.~J. van~der Laan.
\newblock One-step targeted maximum likelihood estimation for targeting cause-specific absolute risks and survival curves.
\newblock \emph{Biometrika}, 111(1):129--145, 2024.
\newblock \url{https://doi.org/10.1093/biomet/asad033}.

\bibitem[Shi et~al.(2019)]{shi2019dragonnet}
C.~Shi, D.~Blei, and V.~Veitch.
\newblock Adapting neural networks for the estimation of treatment effects.
\newblock \emph{Advances in Neural Information Processing Systems}, 32, 2019.
\newblock \url{https://proceedings.neurips.cc/paper/2019/hash/8fb5f8be2aa9d6c64a04e3ab9f63feee-Abstract.html}.

\bibitem[Zheng and van~der Laan(2011)]{zheng2011cvtmle}
W.~Zheng and M.~J. van~der Laan.
\newblock Cross-validated targeted minimum-loss-based estimation.
\newblock In \emph{Targeted Learning: Causal Inference for Observational and Experimental Data}, pages 459--474. Springer, 2011.
\newblock \url{https://doi.org/10.1007/978-1-4419-9782-1_27}.

\bibitem[van~der Laan and Rose(2011)]{vdl2011tl}
M.~J. van~der Laan and S.~Rose.
\newblock \emph{Targeted Learning: Causal Inference for Observational and Experimental Data}.
\newblock Springer, 2011.
\newblock \url{https://doi.org/10.1007/978-1-4419-9782-1}.

\bibitem[van~der Laan et~al.(2023)]{vdl2023adml}
L.~van~der Laan, M.~Carone, A.~Luedtke, and M.~J. van~der Laan.
\newblock Adaptive debiased machine learning using data-driven model selection techniques.
\newblock \emph{arXiv preprint arXiv:2307.12544}, 2023.
\newblock \url{https://arxiv.org/abs/2307.12544}.

\end{thebibliography}

\end{document}